\documentclass[a4paper,11pt]{article}
\pdfoutput=1

\usepackage{jheppub}

\usepackage{graphicx,subfigure,multirow}
\usepackage{color}
\usepackage[usenames,dvipsnames]{xcolor}

\makeatletter

\newcommand{\fmslash}[2][0mu]{%
  \mathchoice
    {\fmsl@sh\displaystyle{#1}{#2}}%
    {\fmsl@sh\textstyle{#1}{#2}}%
    {\fmsl@sh\scriptstyle{#1}{#2}}%
    {\fmsl@sh\scriptscriptstyle{#1}{#2}}}
\newcommand{\fmsl@sh}[3]{%
  \m@th\ooalign{$\hfil#1\mkern#2/\hfil$\crcr$#1#3$}}
\makeatother

\newcommand{\lsim}{{\;\raise0.3ex\hbox{$<$\kern-0.75em\raise-1.1ex\hbox{$\sim$}}\;}}
\newcommand{\gsim}{{\;\raise0.3ex\hbox{$>$\kern-0.75em\raise-1.1ex\hbox{$\sim$}}\;}}

\newcommand{\beq}{\begin{equation}}
\newcommand{\eeq}{\end{equation}}
\newcommand{\bea}{\begin{eqnarray}}
\newcommand{\eea}{\end{eqnarray}}
\mathchardef\minus="002D

\usepackage{color}

\makeatletter
\newcommand\footnoteref[1]{\protected@xdef\@thefnmark{\ref{#1}}\@footnotemark}
\makeatother

\title{Dark Matter ``Transporting'' Mechanism Explaining Positron Excesses}

\author[a]{Doojin Kim,}
\author[b]{Jong-Chul Park,\footnote{\label{note1}Corresponding author.}}
\author[c,d]{Seodong Shin\footnoteref{note1}}

\affiliation[a]{Theory Division, CERN, CH-1211 Geneva 23, Switzerland}
\affiliation[b]{Department of Physics, Chungnam National University, Daejeon 34134, Republic of Korea}
\affiliation[c]{Department of Physics \& IPAP, Yonsei University, Seoul 03722, Republic of Korea}
\affiliation[d]{Enrico Fermi Institute, University of Chicago, Chicago, Illinois 60637, USA}
\emailAdd{doojin.kim@cern.ch, jcpark@cnu.ac.kr, shinseod@indiana.edu}

\preprint{
CERN-TH-2017-036\\
}

\abstract{
We propose a {\it novel} mechanism to explain the positron excesses, which are observed by satellite-based telescopes including PAMELA and AMS-02, in dark matter (DM) scenarios.
The {\it novelty} behind the proposal is that it makes direct use of DM {\it around the Galactic Center} where DM populates most densely, allowing us to avoid tensions from cosmological and astrophysical measurements.
The key ingredients of this mechanism include DM annihilation into {\it un}stable states with a very long laboratory-frame life time and their ``retarded'' decay near the Earth to electron-positron pair(s) possibly with other (in)visible particles.
We argue that this sort of explanation is {\it not} in conflict with relevant constraints from big bang nucleosynthesis and cosmic microwave background.
Regarding the resultant positron spectrum, we provide a generalized source term in the associated diffusion equation, which can be readily applicable to any type of two-``stage'' DM scenarios wherein production of Standard Model particles occurs at completely different places from those of DM annihilation.
We then conduct a data analysis with the recent AMS-02 data to validate our proposal.
}

\begin{document}

\maketitle

\section{Introduction}
Very recently, the AMS-02 Collaboration has released new results from a set of data accumulated for past five years~\cite{AMSrecent}.
The reported positron flux and fraction clearly exhibit a rise from $\sim 10$ GeV above the rate expected from cosmic-ray collisions, which is consistent with previous results by PAMELA~\cite{Adriani:2008zr, Adriani:2013uda}, Fermi-LAT~\cite{FermiLAT:2011ab}, and AMS-02~\cite{Aguilar:2013qda, Accardo:2014lma}. While astrophysical sources such as pulsars~\cite{Hooper:2008kg, Delahaye:2014osa} or supernova remnants~\cite{Hu:2009bc} would eventually explain the excess, dark matter (DM) interpretations have received ceaseless attention as an orthogonal attempt.

The positron flux $\Phi$ from annihilating DM is essentially determined by the DM density $\rho$ and the velocity-averaged annihilation cross section $\langle \sigma v \rangle$:
\begin{align}
\Phi \propto \rho^2\langle \sigma v \rangle\,. \label{eq:flux}
\end{align}
We remark that the positrons produced within $\sim$1 kpc from the Earth dominantly contribute to the observed flux.
The local density and the typical annihilation cross section of the thermal relic DM, however, predict a much smaller flux than the observed.
Therefore, it has been a major challenge to identify an adequate DM source to supply the measured positron flux.

Several ideas have been proposed to resolve this issue.
The first set of attempts is to secure enough flux by enhancing the annihilation cross section today, i.e., the second component in eq.~\eqref{eq:flux}, compared to what is required by the standard thermal production of DM, with the current relic abundance being the same.
Example mechanisms include Sommerfeld enhancement~\cite{Hisano:2002fk, ArkaniHamed:2008qn, Chun:2008by, Rothstein:2009pm} and relaxation of the thermal relic relation via late decays of DM partners~\cite{Fairbairn:2008fb}.
However, any moving charged particles radiate photons so that the gamma-ray data from Milky Way satellite galaxies~\cite{Ackermann:2015zua, Ahnen:2016qkx, Fermi-LAT:2016uux} often sets quite stringent limits on the allowed DM annihilation cross section, while they may be relaxed by dispersing the positron production zone with a long-lived intermediary state disintegrating to positrons~\cite{Rothstein:2009pm}.

The second set of attempts is to increase the local DM density itself, i.e., the first component in eq.~\eqref{eq:flux}, by a clumpy DM distribution.
$\Lambda$CDM $N$-body simulation studies, however, suggest that the flux enhancement by the local clumpy DM distribution may not suffice to satisfy the observed data~\cite{Lavalle:1900wn,Cline:2010ag}.
Due to these difficulties in explaining the large positron excess by annihilating DM, the other class of attempts utilizes decaying DM with a long life time of $\sim 10^{26}$ seconds~\cite{Chen:2008yi}.
However, typical decaying DM models favored by the positron excesses are also excluded by or in tension with Fermi-LAT gamma-ray observations, depending on decay channels and modeling of astrophysical foregrounds and backgrounds~\cite{Ando:2015qda, Massari:2015xea, Liu:2016ngs}.
Given the drawbacks of previous trials, in this paper, we propose a {\it novel}, alternative mechanism to invoke a sufficient positron flux by taking {\it annihilating/decaying DM around the Galactic Center} (GC), where DM densely populates, in the context of non-minimal dark-sector scenarios.

To present the main ideas efficiently, the paper is organized as follows. In section~\ref{sec:mechanism}, we elaborate the mechanism proposed here, followed by a detailed comparison of the mechanism with others. We then discuss how the electron or positron created in the scenario under consideration propagates in the galaxy, in section~\ref{sec:propagation}. Section~\ref{sec:data} contains our main result to reproduce the positron excess reported by the AMS-02 Collaboration with the formulation developed in the preceding section. Finally, we summarize and conclude the study performed in this paper in section~\ref{sec:conclusion}.

\section{Mechanism \label{sec:mechanism}}

\subsection{Main idea}
\begin{figure}[t]
\centering
\includegraphics[width=0.62\linewidth]{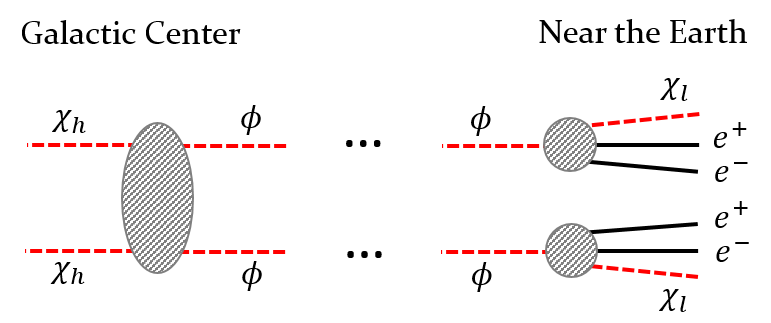}
\caption{\label{fig:scenario} A benchmark scenario for the positron excesses.}
\end{figure}

Our mechanism is predicated upon a non-conventional situation, wherein DM particles annihilate or decay
{\it not} promptly to leptonic final states {\it but} to {\it un}stable particles around the GC.
We assume that this unstable particle has a sufficiently long life time to propagate a large enough distance from the GC and decay to electron-positron pair(s) potentially with other particles near the Earth.
Figure~\ref{fig:scenario} delineates our benchmark scenario for the positron excesses: a pair of (heavier) DM particles $\chi_h$ annihilate to a pair of unstable (possibly dark-sector) states $\psi$ each of which subsequently disintegrates to an electron, a positron, and a (lighter) DM particle $\chi_l$, in the vicinity of the Earth, via a three-body decay process.
One may view the role of $\psi$ as proxy for DM in that it ``transports'' DM at the GC, which {\it effectively} enhances $\rho$ near the Earth and in turn, the positron flux $\Phi$ itself with $\langle \sigma v \rangle_{\chi_h \chi_h \to \psi \psi} \sim 10^{-26}\,{\rm cm}^3 / {\rm s}$.
We emphasize that the scenario under consideration is similar to what arises in typical boosted DM (BDM) scenarios~\cite{Agashe:2014yua, Berger:2014sqa, Kong:2014mia, Alhazmi:2016qcs} based on the {\it assisted freeze-out} mechanism~\cite{Belanger:2011ww} (modulo the heavier dark-sector state like in {\it inelastic} BDM scenarios~\cite{Kim:2016zjx, Giudice:2017zke}), in which $\chi_h$ is the dominant relic component while $\chi_l$ typically comprises of $\ll 1\%$.

The survival rate by which a $\psi$ arrives near the Earth without breaking apart and its subsequent decay are dictated by the decay width $\Gamma_\psi$ and the Lorentz boost factor $\gamma_\psi$ of $\psi$.
We remark that the Earth is quite distant from the GC, implying that typical sizes of the dilated life time $\tau_\psi \gamma_\psi$ should be as large as $\sim 8\times10^{11}$ seconds to travel about 8 kpc.
Assuming that the heavier DM species $\chi_h$ is non-relativistic, we see that the $\psi$ mass $m_{\psi}$ has a simple relation with the $\chi_h$ mass $m_{\chi_h}$ as $m_{\chi_h} = \gamma_\psi m_{\psi}$.
Stringent constraints~\cite{Poulin:2015opa, Poulin:2016anj} on the long-lived particle $\psi$, which stem from big bang nucleosynthesis (BBN) and cosmic microwave background (CMB), enforces us to consider the scenario with a large boost factor $\gamma_\psi$.
Therefore, we are able to treat both $\psi$ and $\chi_l$ effectively massless compared to $m_{\chi_h}$.

Obviously, under this mass hierarchy, $m_{\chi_h}$ governs the overall scale of the positron energy spectrum.
Adopting the nominal positron spectrum by AMS-02~\cite{AMSrecent} which drops at high energies, we expect $m_{\chi_h}$ to be in the range of $\sim1$ TeV.
On the other hand, $m_\psi$ should be greater than the sum of positron and electron masses, i.e., $\sim$1 MeV.
We then find that the value of $\gamma_\psi$ reaches at most a few $10^6$, implying that its minimum life time should be greater than a few $10^5$ seconds in order for $\psi$ produced around the GC to travel close enough to the Earth.
As mentioned above, due to this long life time of $\psi$, some cosmological constraints come into play.
Above all, since $\psi$ decays to charged particles after BBN, they would affect the evolution of nuclei fractions in the history of the Universe.
The constraints from BBN and CMB depend on the energy density of $\psi$ at the early Universe ($\rho_\psi$) times the fraction of decay energy going to stable photon and $e^\pm$, as displayed in Figure~5 of Ref.~\cite{Poulin:2016anj}.
For $\rho_\psi \sim 10^{-2} - 10^{-5}$ relative to the DM relic $\rho_{\rm DM}$ (mostly $\chi_h$ here), the life time of $\psi$ is limited as $\tau_\psi \lesssim 10^6 - 10^8$ seconds by BBN constraints, under the assumption that 100\% of the decay energy is deposited to stable photon and $e^\pm$.
If $\rho_\psi$ is in-between $10^{-5}$ and $10^{-11}$ of $\rho_{\rm DM}$, the existence of $\psi$ is constrained not by BBN but by CMB, requiring $\tau_\psi \lesssim 10^{12}$ seconds.
For $\rho_\psi \lesssim 10^{-11}\cdot\rho_{\rm DM}$, there are no constraints even from the CMB observation.

We remark that as a spin-off, the existence of a relativistic long-lived particle $\psi$ allows us to evade the observational $\gamma$-ray bounds from the GC and dwarf spheroidal galaxies, as the production of charged Standard Model particles occurs far away from those regions after $\psi$ is sufficiently dispersed.
Furthermore, the charged particles produced outside the galactic cylinder never re-enter because of the large boost factor $\gamma_\psi$. Interestingly, this kind of set-up makes it possible to avoid the bound from the $\gamma$-ray flux by inverse Compton scattering (ICS) as well. Our choice of $\langle \sigma v \rangle \sim 10^{-26}\,{\rm cm}^3 / {\rm s}$ and $m_{\chi_h} \sim 1$ TeV is apparently safe from the bound reported by the Fermi-LAT Collaboration~\cite{Ackermann:2012rg} in the sense that our process is topologically similar to the one that the ordinary DM pair annihilates to a pair of muons, followed by their three-body decay.

Another important issue in realizing our mechanism is to secure an enough $\psi$ flux near the Earth for explaining the positron bump reported by AMS-02.
In other words, a sufficient amount of DM should be guaranteed near the GC.
While ordinary cuspy DM halo profiles in the market yields an $\mathcal{O}(1)$ enhancement in the positron flux, we remind that the DM density near the galactic core still comes with a huge uncertainty irrespective of the choice of DM profiles~\cite{Cirelli:2010xx, Kim:2016csm}.
Reflecting this uncertainness, we shall treat the core size and the DM density inside the core as free parameters in our data analysis.

\subsection{Comparison with other mechanisms}
It is rather informative to highlight novel features in the mechanism, which are deeply related to ways of ameliorating potential issues and challenges in other existing mechanisms.
For the mechanisms to explain the positron excess via DM annihilation, they are essentially categorized according to how to boost up the positron flux to accommodate the amount of observed data.
As mentioned in the introductory section, our proposal here can fall into the same category as the ones increasing the local DM density.
However, as elaborated in the previous section, it is {\it effectively} achieved by making use of the big DM clump around the GC through the ``retarded'' decay (near the Earth) of an unstable intermediary dark-sector state, and the {\it relativistic} nature of the intermediary particle $\psi$ plays a crucial role in getting around various cosmological and astrophysical bounds.

On the other hand, the mechanisms elevating the velocity-averaged DM annihilation cross section $\langle \sigma v \rangle$ at the current universe belong to a different category.
They are usually constrained by the $\gamma$-ray observations coming from the GC or dwarf spheroidal galaxies.
Indeed, a possible way to alleviate this constraint is to introduce a long-lived dark-sector state to disperse the positron production zone~\cite{Rothstein:2009pm} in association with a similar event topology to ours.
However, this is still in the category of inflating $\langle \sigma v \rangle$ (of a DM pair into a pair of long-lived particles) by Sommerfeld enhancement, i.e., the key idea explaining the large positron spectrum is totally different from our mechanism in this paper.
More quantitatively, the required enhancement in Ref.~\cite{Rothstein:2009pm} is as large as $\mathcal{O}(10^3$) in order to keep $\langle \sigma v \rangle_{\textrm{freeze-out}} \sim 10^{-26} {\rm cm}^3/{\rm s}$.
This sort of Sommerfeld enhancement solution potentially leads to several phenomenological problems including a too large contribution to the diffuse extragalactic
$\gamma$-ray background~\cite{Kamionkowski:2008gj}, which is later supported by a Fermi-LAT observation~\cite{Ackermann:2012rg}, and additional model building for the new attractive force as already addressed in Ref.~\cite{Rothstein:2009pm}.
Note that it is highly non-trivial to get such large Sommerfeld enhancement if the force carrier is much lighter than the DM, hence produced relativistically~\cite{Feng:2009hw, Feng:2010zp}.
Therefore, it is hard to contain a relativistically produced long-lived particle (i.e., $\psi$ in our notation) in the mechanism in Ref.~\cite{Rothstein:2009pm}.
Moreover, the non-relativistically produced long-lived particles may suffer from completely different cosmological and astrophysical issues:
for example, an additional mechanism to dilute the number density of the long-lived particle should be supplemented to avoid the BBN bounds~\cite{Poulin:2015opa, Poulin:2016anj} because its mass is nearly equal to that of DM.

\section{Electron and Positron Propagation \label{sec:propagation}}

In this section, we discuss the spread of electrons and positions in the interstellar medium.
Once electron-positron pairs are created by the $\psi$ decay, they should propagate to the Earth according to the following diffusion equation:
\bea
 \frac{\partial }{\partial t}f(\vec{x},E)  -\vec{\nabla}\cdot\left[K(\vec{x},E)\vec{\nabla} f(\vec{x},E)\right] -\frac{\partial}{\partial E}\left[ b(\vec{x},E)f(\vec{x},E)\right]=Q(\vec{x},E)\,.
\label{eq:diffusion}
\eea
Here $f(\vec{x},E)$ denotes the electron and positron differential number density with energy $E$ at $\vec{x}$, from which the differential flux of $e^{\pm}$ is evaluated as
\bea
\frac{d}{dE}\Phi(\vec{x},E)=\frac{v}{4\pi}f(\vec{x},E)\,,
\eea
where $v$, the velocity of $e^{\pm}$, is essentially the same as the speed of light.
$K(\vec{x},E)$ and $b(\vec{x},E)$ describe the diffusion and energy-loss rates by external electromagnetic activities during propagation, respectively.
We point out that in the standard DM scenario the source term $Q(\vec{x},E)$ is associated with the DM density at the same position because DM promptly annihilates to electrons and positrons in the final state.
However, in our setup, $\psi$ is ultra-relativistic and propagates very far without decaying to an $e^+e^-$ pair, motivating us to modify $Q(\vec{x},E)$ by carefully incorporating non-conventional aspects.

\begin{figure}
\centering
\includegraphics[width=0.67\linewidth]{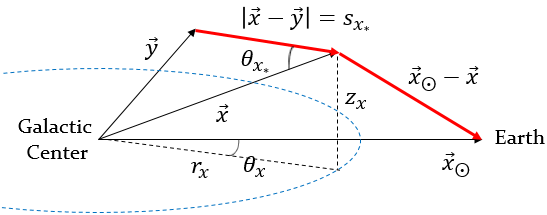}
\caption{\label{fig:config}
Schematic situation from production and decay of $\psi$ at $\vec{y}$ and $\vec{x}$ to propagation of $e^{\pm}$.
All position vectors are defined with respect to the GC as the origin.
$s_{x_*}$ and $\theta_{x_*}$ are measured with respect to $\vec{x}$.
}
\end{figure}

Figure~\ref{fig:config} schematically shows the situation at hand: the production and the decay of $\psi$ take place at $\vec{y}$ and $\vec{x}$, respectively, and $e^{\pm}$ from the $\psi$ decay transports to the solar system $\vec{x}_{\odot}$.
The source at $\vec{x}$ can be described by the $\psi$ decay:
\begin{align}
Q(\vec{x},E)=n_{\psi}(\vec{x})\,\Gamma_\psi^{\textrm{lab}}\,\frac{dN}{dE}\,,
\end{align}
where $dN/dE$ represents the electron or positron (injection) energy spectrum at $\vec{x}$
measured in the laboratory frame.
The laboratory-frame $\psi$ decay rate $\Gamma_\psi^{\textrm{lab}}$ $(=1/\tau_\psi^{\textrm{lab}})$ is given by the ratio of the rest-frame decay rate $\Gamma_\psi$ $(=1/\tau_\psi)$ to the $\psi$ boost factor $\gamma_\psi$:
\begin{align}
\Gamma_\psi^{\textrm{lab}} \equiv \frac{\Gamma_\psi}{\gamma_\psi}\,.
\end{align}
As usual, the $\psi$ flux $\Phi_\psi(\vec{x})$ relates the $\psi$ density $n_\psi(\vec{x})$ and its speed $v_\psi$ as
\begin{align} \label{eq:fluxphi}
\Phi_\psi=n_\psi\cdot v_\psi\,,
\end{align}
where $v_\psi$ is simply given by the speed of light $c$ for relativistic $\psi$.
Given the assumption that $\psi$ is non-diffusive, we formulate $\Phi_\psi$ in an analogous manner to the case of the DM annihilation to a photon pair as follows:
\bea
\frac{d\Phi_\psi(\vec{x})}{d\Omega_{x_*}dE_\psi}
=\left(\frac{1}{2}\right)\cdot \frac{1}{4\pi}&&\hspace{-0.2cm}\int_{\textrm{l.o.s}} ds_{x_*}\frac{n_{\chi_h}^2(\vec{y})}{2} \langle\sigma v\rangle_{\chi_h\chi_h\rightarrow \psi\psi} \,e^{-\frac{|\vec{x}-\vec{y}|}{c}
\Gamma_\psi^{\textrm{lab}}}
\frac{dN_\psi}{dE_\psi}~, \label{eq:modifiedQ}
\eea
where the additional factor of two in the parentheses is available when $\chi_h$ and $\bar{\chi}_h$ are distinguishable.
Here $\textrm{l.o.s}$ implies a line-of-sight integral along the $(\vec{x}-\vec{y})$ direction with the coordinates $ds_{x_*}$ and $d\Omega_{x_*}$ defined relative to $\vec{x}$, and therefore we find
\bea
|\vec{x}-\vec{y}|=s_{x_*}\,.
\eea
The exponential factor describes the survival rate of $\psi$ from $\vec{y}$ to $\vec{x}$ without disintegrating into $e^+e^-$.
$dN_\psi/dE_\psi$ represents the (injection) energy spectrum of $\psi$, which is simply given by
\bea
\frac{dN_\psi}{dE_\psi}=2\cdot\delta (E_\psi -m_{\chi_h})\,,
\eea
since we assume $\chi_h$ non-relativistic.\footnote{Note that the $\psi$ flux in eq.~\eqref{eq:modifiedQ} straightforwardly applies to decaying DM models by replacing $n_{\chi_h}^2 (\vec{y})/2$ and $\langle\sigma v\rangle_{\chi_h\chi_h \rightarrow \psi\psi}$ by $n_{\chi_h}(\vec{y})$ and $\Gamma_{\chi_h\rightarrow \psi\psi}$, respectively.}

A couple of comments are in order.
First, as we shall see shortly, the $\chi_h$ number density depends only on $y \equiv |\vec{y}|$.
We then express $y$ in terms of $s_{x_*}$ and $\cos\theta_{x_*}$ as follows:
\bea
y^2 = r_x^2 + z_x^2+s_{x_*}^2-2\sqrt{r_x^2 + z_x^2}\,s_{x_*}\cos\theta_{x_*}\,,
\eea
where $x^2=r_x^2+z_x^2$ in the cylindrical coordinate relative to the GC (see also Figure~\ref{fig:config}).
Second, one may argue that the expected positron spectrum in the DM scenarios of interest would come with a significant anisotropy since particle $\psi$, the immediate source of $e^{\pm}$ is highly relativistic and its decay is substantially delayed.
However, the fact that the source point fairly far away is not resolvable in the point-like source interpretation (e.g., pulsars) for the positron excess~\cite{AMSrecent} suggests that
the positron flux in our case appears as almost isotropic at the current level of the data.\footnote{Recently, a concrete analysis on the anisotropy of the positron from our scenario is announced in Ref.~~\cite{Chu:2017vao}.}
Future experiments equipped with better sensitivity would observe the degree of anisotropy (in the direction of the GC), which is anticipated under our proposed scenario to explain the positron excesses.

\section{Data Analysis \label{sec:data}}
Equipped with the formalism developed thus far, we are now in the position to conduct a data analysis with the AMS-02 data in order to test the validity of our proposed mechanism.
We basically vary the mass parameters for particles $\chi_h$, $\psi$, and $\chi_l$ in order to find out the best set of parameter values.

\subsection{DM density profile and injection spectrum}

\begin{figure*}
\centering
\includegraphics[width=0.5\linewidth]{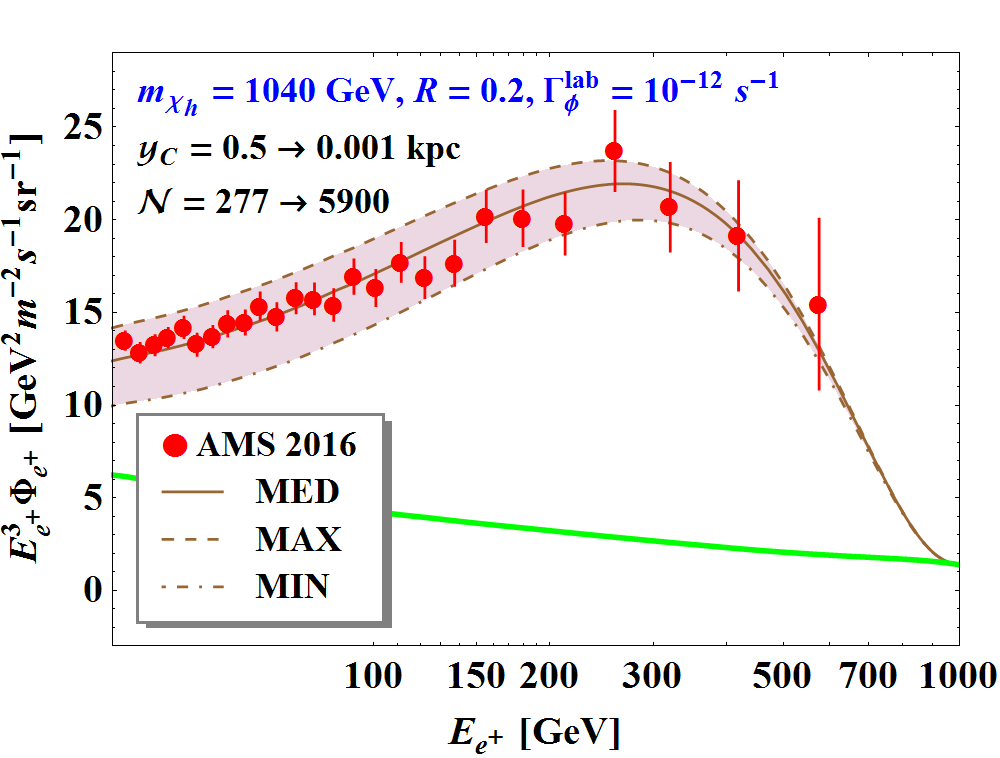} \\
\includegraphics[width=0.487\linewidth]{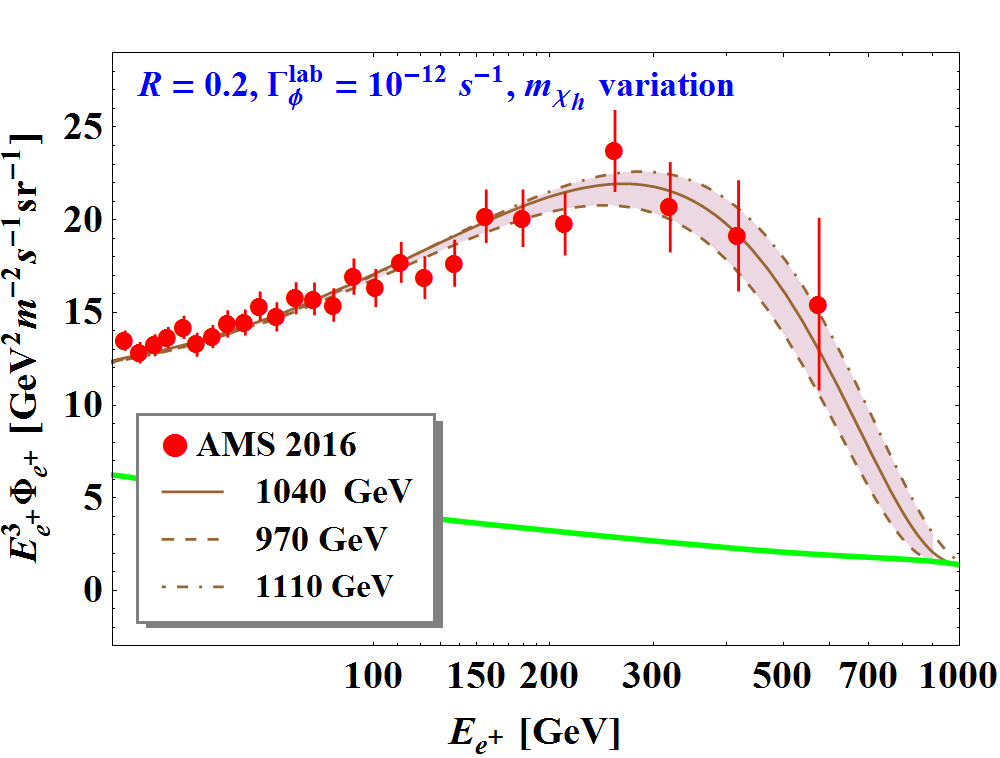}\hspace{0.28cm}
\includegraphics[width=0.487\linewidth]{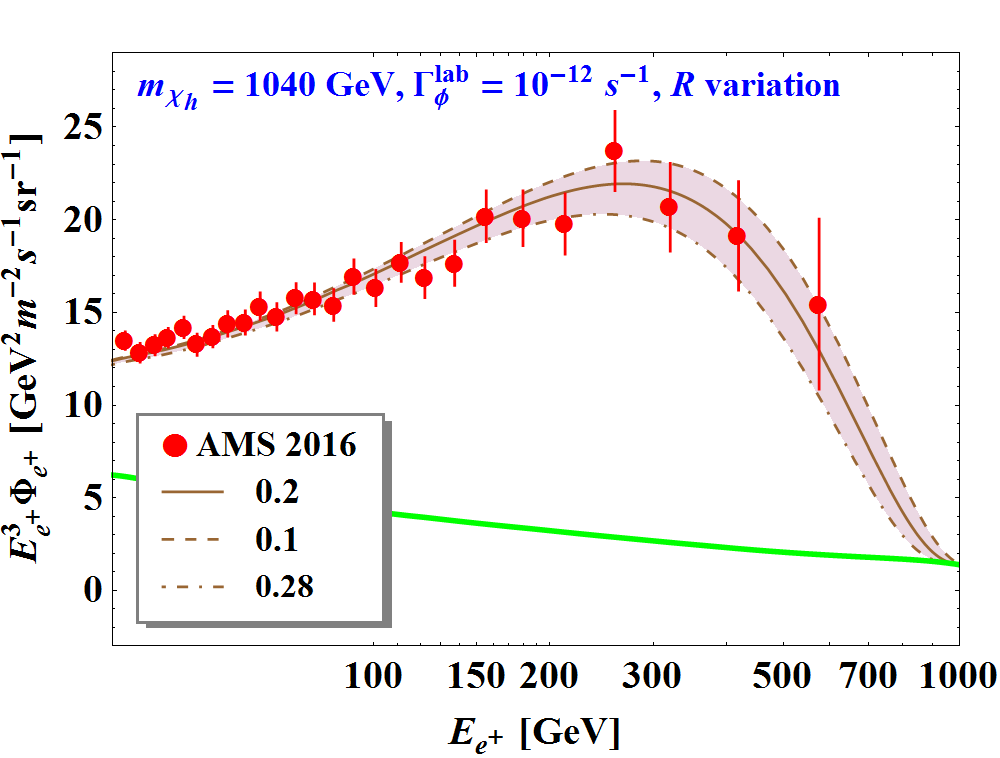}
\caption{\label{fig:spectrum} The fit results of the scenario in Figure~\ref{fig:scenario} into the positron spectrum reported by the AMS-02 Collaboration~\cite{AMSrecent}.
The fits are performed with 26 data points within 35 -- 600 GeV.
Upper panel: the best-fit with the variations of $y_C$, $\mathcal{N}$, and the galactic cylinder model.
The variation in $y_C$ and $\mathcal{N}$ is performed under the MED model, while the best fit is retained, i.e., the choice of $(y_C,\mathcal{N})=(0.5 \hbox{ kpc}, 277)$ yields the same minimum $\chi^2$ value as the choice of $(y_C,\mathcal{N})=(0.001 \hbox{ kpc}, 5900)$ does.
Lower-left and lower-right panels: the fit sensitivities to $m_{\chi_h}$ (lower-left) and $R$ (lower-right) with the other fit parameters set to be their respective best values. $R$ stands for the ratio of the lighter DM mass to the unstable state mass, i.e., $R \equiv m_{\chi_l}/m_\psi$. 
For both cases, we choose $y_C=0.5$ kpc, $\mathcal{N}=277$, and the MED model. }
\end{figure*}

As mentioned earlier, we implement the uncertainness of the DM density around the GC, employing a $\chi_h$ profile wherein $n_{\chi_h}$ is enhanced nearby the GC compared to usual cuspy profiles.
A simple example is given as
\bea
\rho_{\chi_h}(y) =\left\{
\begin{array}{l l}
 \rho_0\, \frac{(y/y_s)^{-1}}{(1+y/y_s)^{2}} \equiv \rho_{\rm NFW} (y) & {\rm for}~~ y \ge y_C \\ [4mm]
\mathcal N \times \rho_{\rm NFW} (y_C)  & {\rm for}~~ y < y_C
\end{array} \right.,
\label{eq:profile}
\eea
where $y_C$ and $\mathcal{N}$ are fit parameters responsible for the core size and the density scale factor, respectively.
A scale radius $y_s$ is set to be 20 kpc, while $\rho_0$ is chosen in such a way that the local DM density at $r_\odot \simeq 8.33$ kpc is $\rho_\odot \simeq 0.4 ~{\rm GeV}/{\rm cm}^3$.
Our toy profile implies that the $\chi_h$ density outside $y_C$ simply follows the Navarro-Frenk-White (NFW) halo profile~\cite{Navarro:1995iw, Navarro:1996gj}, while there exists a large amount of $\chi_h$ inside $y_C$ with a flat and central profile.
As we will see shortly, an excellent fit to the AMS-02 data arises with $\mathcal N = 277\,(5900)$ for $y_C = 0.5\,(10^{-3})$ kpc (see the upper panel in Figure~\ref{fig:spectrum}).
We emphasize that the total amount of DM for our toy profile within 60 kpc around the GC remains almost the same as the NFW profile case due to the small volume defined by the enhanced core radius.
Furthermore, our reference profiles were examined by a recent work~\cite{Chu:2017vao} in the context of the adiabatic contraction effect arising from the formation of the black hole around the GC; the authors in Ref.~\cite{Chu:2017vao} found that the adiabatic contraction can marginally induce a large DM spike, which is desired to explain the positron excess in our mechanism, under reasonable assumptions and circumstances.\footnote{A potential issue might arise in association with a recent study on estimating the amount of DM in the galactic bulge-bar region~\cite{bulge}. We have explicitly checked that the additional amount of DM by the latter choice of best-fit parameters (i.e., $\mathcal N = 5900, y_C = 1$\,pc) is well below their estimation. By contract, the former (i.e., $\mathcal N= 277, y_C = 0.5$\,kpc) is in marginal tension, which will be resolved in future analyses. }

In order to obtain the injection spectrum $dN/dE$, we take a three-body decay of $\psi$ by phase space.
The decay may involve a non-trivial matrix element, potentially distorting the overall shape.
However, it has been shown that such an effect upon the three-body decay is indeed subleading in well-motivated new physics scenarios~\cite{Cho:2012er}.
In addition, given the typical size of $\gamma_\psi$, phase-space decay can be a good approximation.
In the massless limit of the electron/positron, the unit-normalized injection spectrum has a form of
\bea
\frac{dN}{dE}&=& \frac{1}{2\gamma_\psi\beta_\psi} \left[\frac{m_\psi^4-m_{\chi_l}^2}{8m_\psi}+\frac{m_{\chi_l}^2m_\psi}{2}\log\left(\frac{m_{\chi_l}}{m_\psi} \right) \right]^{-1}
\nonumber \\
&\times &
\left[  m_\psi(E^+-E^-)+\frac{m_{\chi_l}^2}{2}\log\left(\frac{m_\psi-2E^+}{m_\psi-2E^-}\right) \right]\,, \label{eq:injection}
\eea
where $E^{\pm}$ are defined as follows:
\bea
E^+ &=& \min\left[\frac{E}{\gamma_\psi(1-\beta_\psi)},\frac{m_\psi^2-m_{\chi_l}^2}{2m_\psi}\right]\,, \\
E^- &=& \frac{E}{\gamma_\psi(1+\beta_\psi)}\,.
\eea
Note that $E$ spans 0 to $\gamma_\psi(1+\beta_\psi)(m_\psi^2-m_{\chi_l}^2)/(2m_\psi)$.

Although in our actual data analysis we use the full form in eq.~\eqref{eq:injection}, it is instructive to examine the injection spectrum in a phenomenologically well-motivated mass hierarchy, i.e., $m_{\chi_h} \gg m_\psi \gg m_{\chi_l}$, in order to check its effective dependence over the mass parameters.
We find that a simple algebra can further simplify the expression in eq.~\eqref{eq:injection} to
\bea
\frac{dN}{dE}\sim \frac{m_{\chi_h}-E}{m_{\chi_h}}+R^2\log\left[R^2\left(\frac{m_{\chi_h}-E}{m_{\chi_h}} \right)^{-1} \right],
\eea
with $R=m_{\chi_l}/m_\psi$. In this limit, the positron spectrum becomes sensitive only to $m_{\chi_h}$ and $R$ among three masses.

\subsection{Data fit strategy and results}
For obtaining the actual positron spectrum, we simply apply the results given in PPPC4DM~\cite{Cirelli:2010xx} rather than directly solve the diffusion equation, eq.~\eqref{eq:diffusion}.
In general, the (induced) density profile of $\psi$ differs from all of the DM halo profiles implemented in the PPPC4DM package.
However, for explaining positron excesses, the relevant features of halo functions are restricted mostly near the Earth where the halo profiles barely show variation.
Thus, we can use any conventional profiles implemented in the package with the overall scale factor normalized to the (induced) $\psi$ density near the Earth, and then adopt the formalism for the decay scenario as $e^{\pm}$'s are from the $\psi$ decay.
To reduce the number of fit parameters, we fix $\langle \sigma v\rangle_{\chi_h \chi_h \rightarrow \psi \psi}$ and $\Gamma_\psi^{{\rm lab}}$ to be $3\times 10^{-26}$ cm$^3$s$^{-1}$ and $10^{-12}$ s$^{-1}$, respectively, and take the third magnetic field model in PPPC4DM throughout our data analysis.\footnote{We have tried the other magnetic field models, but observed no significant differences.}
The conventional magnetic field profile in the galaxy is given by \cite{Strong:1998fr}
\bea
B(r,z)=B_0 \exp \left[-\frac{r-r_\odot}{r_B}-\frac{|z|}{z_B} \right]\,,
\eea
and the third model takes $B_0=9.5\, \mu$G, $r_B=30$ kpc, and $z_B=4$ kpc.
Also, we assume that $\chi_h$ is self-conjugate and $\psi$ exclusively decays into the electron-positron pair along with $\chi_l$ for the sake of simplicity.

We now demonstrate the best-fit spectrum for the nominal positron excess announced by AMS-02 in the upper panel of Figure~\ref{fig:spectrum} where the dependence on the propagation parameters appears as a shaded band for the well-established MIN, MED, and MAX models~\cite{Donato:2003xg}.
We also dial $y_C$ and $\mathcal{N}$ as well while keeping the best fit (i.e., the minimum $\chi^2$ value in the fit) with the MED model, and observe that the required value of $\mathcal N$ (relatively) mildly increases as $y_C$ decreases.
To develop our intuition on the mass spectrum dependence of the fit, we vary $m_{\chi_h}$ and $R$ with the other parameters fixed to those in the best fit.
The lower-left panel of Figure~\ref{fig:spectrum} shows the former variation together with $y_C=0.5$ kpc, $\mathcal{N}=277$, and the MED model. 
We find its best-fit value with the 90\% confidence interval to be $m_{\chi_h}=1040\pm70$ GeV. 
As $m_{\chi_h}$ determines the maximum positron energy, we clearly see the corresponding shift in three curves.
The lower-right panel of Figure~\ref{fig:spectrum}, on the other hand, shows the latter variation again with $y_C=0.5$ kpc, $\mathcal{N}=277$, and the MED model. 
The corresponding best-fit value with the 90\% confidence interval is $R=0.20^{+0.08}_{-0.01}$.
We note that a larger (smaller) value of $R$ implies that a smaller (larger) fraction of the $\psi$ decay energy is carried away by the positron. So, the resulting spectrum becomes softer (harder) as $R$ increases (decreases), which is apparently respected in our fit results.
We have also explicitly checked that $\sim50\%$ variation in the best-fit $\Gamma_\psi^{{\rm lab}}$ still keeps $\sim90\%$ of the flux.

Finally, let us discuss the choice of $m_{\chi_l}$ and $m_\psi$, although only the ratio $R$ between them enters in determining the shape of the positron spectrum.
One can avoid the constraints from BBN~\cite{Poulin:2016anj} up to $\tau_\psi\simeq10^{12}$ seconds when the CMB bounds come into play, as far as the following relation holds:
\bea
\frac{2}{3}\cdot \frac{\rho_\psi}{\rho_{\rm DM}} = \frac{2}{3}\cdot \frac{m_\psi n_\psi}{m_{\rm DM} n_{\rm DM}} \lesssim 2 \times 10^{-5}\,.
\eea
In the scenarios where the number densities of $\chi_h$ and $\psi$ are similar at the early Universe ($n_{\chi_h} \simeq n_\psi$),
we can simply read off $m_\psi / m_{\chi_h} = \gamma_\psi^{-1}\lesssim 3 \times 10^{-5}$ by assuming that $\chi_h$ is the dominant DM relic.
Given that $\Gamma_\psi^{\rm lab} \sim 10^{-12}\, {\rm s}^{-1}$ and $m_{\chi_h} \sim$ 1 TeV, therefore, $m_\psi \lesssim 30$ MeV can be a proper parameter choice realizing our novel mechanism for positron excesses.
On the other hand, other scenarios of $n_{\chi_h} \gg n_\psi$ are allowed, given that the $\psi$ number density can be reduced by its pair annihilation, e.g., $\psi \psi \to \chi_l \chi_l$.
The exact number densities of $\chi_h$, $\psi$, and $\chi_l$ can be obtained by solving the coupled Boltzmann equations similarly done in some scenarios~\cite{Belanger:2011ww, Bandyopadhyay:2011qm, Dror:2016rxc, Okawa:2016wrr}.
While we leave the detailed calculation for future~\cite{futurework}, our (rough) assessment finds that $\rho_\psi/\rho_{\rm DM} \lesssim 10^{-5}$ for $m_{\chi_h} = 1$ TeV, $m_\psi = 0.5$ GeV, and $m_{\chi_l} = 0.1$ GeV in a dark U(1)$_X$ scenario.
Note again that this parameter choice provides the best fit as displayed in the left panel of Figure~\ref{fig:spectrum}.

\section{Conclusions \label{sec:conclusion}}

In this paper, we provided a {\it new} mechanism which can possibly explain the current and future positron excesses in terms of {\it annihilation/decay} of thermal DM.
A prominent feature of this proposal is that the existence of a very long-lived dark sector particle $\psi$ allows us to make use of the huge DM ($\chi_h$) number density {\it near} the GC to accommodate the data, instead of that in the region close to the Earth where much less DM exists.
A $\chi_h$ pair basically annihilates to a $\psi$ pair near the GC, while the $\psi$ decay (to positron) occurs mostly in the region close to the Earth.
The produced positrons thereby propagate a much shorter distance than 8.33 kpc (between the GC and the Earth) so that a large amount of positron flux can be observed.
This mechanism is in a sharp contrast to other proposals fitting the observed positron spectrum by boosting up the DM annihilation cross section or introducing {\it ad hoc} local DM clumps.
We also argued that quite a broad range of mass spectra are allowed without any severe conflicts with various cosmological and astrophysical observations.
In conclusion, we encourage people to revisit existing DM models or construct new DM models to explain the positron excesses in conjunction with the mechanism suggested here.

\section*{Acknowledgments}
We are particularly grateful to Ji-Haeng Huh for many dedicated discussions and thank Francesco D'Eramo, Roberto Franceschini, Paolo Panci, Joel Primack, and Stefano Profumo for useful and constructive discussions.
We also would like to acknowledge IBS-CTPU for its hospitality and encouraging environment to conceive the idea in this paper.
DK is supported by the Korean Research Foundation (KRF) through the CERN-Korea Fellowship program.
JCP is supported by the National Research Foundation of Korea (NRF-2016R1C1B2015225) and the POSCO Science Fellowship of POSCO TJ Park Foundation. SS is supported by the National Research Foundation of Korea (NRF-2017R1D1A1B03032076).

\end{document}